\documentclass[aps,prc,preprint,superscriptaddress,showpacs,showkeys]{revtex4-1}

\usepackage{graphicx}
\usepackage{amsmath}
\usepackage[left=3cm,top=4cm,right=2cm,bottom=2cm]{geometry}

\begin{document}

\title{Calculating the rates of charmonium dissociation and recombination reactions in heavy-ion collisions using Bateman equation} 
\author{\large Abdulla Abdulsalam}
\affiliation{Department of Physics, King Abdulaziz University, Jeddah, KSA}

\date{\today}

\begin{abstract}

  The charmonium states with their different binding energies and radii dissolve at different temperatures
  of the medium produced in relativistic heavy-ion collisions. Relative yields of charmonium and thus their
  survival have potential to map the properties of Quark Gluon Plasma. In this study, we estimate the combined
  effect of color screening, gluon-induced dissociation and recombination on charmonium production in heavy-ion
  collisions (Pb+Pb ions) at centre of mass energy ($\sqrt{s_{\rm NN}}$) = 5.02 TeV. The rate equations of dissociation
  and recombination are solved separately with a 2-dimensional accelerated expansion of fireball volume. 
  To solve the recombination rate equation, we have used an approach of Bateman solution which
  ensures the dissociation of the recombined charmonium in the QGP medium. The modifications of charmonium
  states are estimated in an expanding QGP with the conditions relevant for Pb+Pb collisions at LHC.

\begin{center}
\end{center}
\end{abstract}



\maketitle

\section{Introduction}\label{Intro}
The relativistic heavy ion collisions paved the way for a comprehensive study of strongly interacting
nuclear matter at high energy density and temperature. According to the theory of Quantum Chromo Dynamics
(QCD), when the temperature of the nuclear matter is increased above a certain value, say
critical temperature $T_C \sim$ 165 MeV~\cite{QGP_Tc}, the matter undergoes a phase transition
to a another state of QCD matter called Quark Gluon Plasma (QGP).
The results from RHIC/LHC experiments \cite{QGP_Tc} are showing signs of formation of this high temperature medium.
One of the most important and interesting signal of QGP is the modification of quarkonium production
and their suppression in the QGP medium. The major cause of the suppression can be Debye Color Screening~\cite{SATZ}
which eventually lead to the dissociation of the states. The ATLAS, CMS and ALICE experiments have
performed quarkonia measurements with Pb+Pb data collected at energies $\sqrt s_{NN}$ = 2.76 TeV and 5.02 TeV.
Measurement of inclusive $J/\psi$ and their nuclear modification factor, $R_{AA}$ computed with 
Pb+Pb data collected at ALICE shows a constant rate of suppression in central collisions, throwing
some hints of recombination/regeneration of charmonia~\cite{ALICE276TeV_Psi,ALICE502TeV_Psi}.
The ATLAS and CMS measurements show suppression of
inclusive, prompt and non-prompt charmonia in central Pb+Pb collisions compared to peripheral collisions
at $\sqrt s_{NN} = 2.76$ TeV and at $\sqrt s_{NN}$ = 5.02 TeV~\cite{ATLAS,ATLAS502,JCMS,JCMS502TeV}.
Since Debye screening length decreases as medium temperature increases, the dissociating pattern of the quarkonia
states depends on their binding energy and radii in the medium.

As an additional source of quarkonium production, recombination/regeneration, would
enhance the number of charmonia in heavy ion collisions, contradicting with the Debye screening
scenario~\cite{Stachel_Reco, Rapp1}. Signs of recombination can be seen in the recent results from
the ALICE Collaboration at the LHC,
which measured a lesser $J/\psi$ suppression than at RHIC~\cite{ALICE276TeV_Psi,ALICE502TeV_Psi},
despite the higher energy collisions. Also, the production of quarkonia in heavy ion collisions are modified
due to non-QGP (non-hot medium) effects such as shadowing~\cite{Vogt} with  
change in nuclear parton distribution function in the small $x$ region compared to that of nucleon~\cite{Muller_Shadow}.

In this paper, we calculate survival probability of charmonium states($J/\psi$, $\psi(2S)$, $\chi_{c}$)
in the deconfined medium of QGP using an extended color screening model of Chu and Matsui \cite{CHU}.
A study was performed on bottomonia suppression using the model and reported in \cite{URatioAbd}.
The present model is improved by adding mechanism like thermal gluon-dissociation, recombination of
charm-quark pairs to have more realistic dynamics of charmonium states in the medium. The competition among
the formation time $\tau_{F}$, medium temperature $T(\tau)$ and lifetime $\tau_{QGP}$ and fireball
expansion decide the trends of the survival probabilities of $\psi$ states in the kinematics of
transverse momentum, $p_{T}$ and centrality, $N_{part}$ (geometry of the heavy-ion collision).
We start by describing the model which provides survival probabilities of $\psi$ states
due to color screening in the medium. Then we describe briefly about the rates of gluon-dissociation
and recombination with the expansion of QGP fireball in transverse and longitudinal direction.
We have used a separate solution for the rate equations of the gluon-dissociation and regeneration reactions
in our calculation which enable us to decouple these processes. The rate equation of gluon-dissociation is solved using
first-order differential equation method and that of recombination is solved by Bateman equation.
In the final section, we present our results from the model calculations followed by a brief discussion
and comparison with experimental data measured at LHC. Details of the other suppression model are available
in the published Ref~\cite{URatioAbd,SuppVineetShukla,Fransua}.

\section{Theoretical Model Formalism}

\subsection{Debye color screening}

According to the model, the QGP is formed at initial entropy density $s_0$ corresponding
to initial temperature $T_0$ at time $\tau_{0}$ which then undergoes an isotropic expansion by Bjorken's
hydrodynamics~\cite{UPsi_Bjork}. The plasma cools to an entropy
density $s_D$ corresponding to the dissociation temperature $T_D$ in time $\tau_{D}$ which is given by 

\begin{equation}\label{bjork}
 \tau_{D} = \tau_0 \left( { s_0 \over s_D} \right) = \tau_0 \left( \frac{T_{0}}{T_{D}} \right)^{3},
\end{equation}
$\tau_0$ is the initial time required for formation of QGP. As long as $\tau_{D}$/$\tau_{F}$ $>$ 1,
the QGP medium will be at high temperature and quarkonium formation will be suppressed. A charm-quark
pair can escape the suppression region, $r_D$ and form $\psi$ if the position at which it is created
satisfies the condition

\begin{equation}\label{taumax}
| {\rm \bf r} + {\tau_{F} {\rm \bf p_{T}} \over M} | > r_D,
\end{equation}
where the suppression region $r$ $<$ $r_D$ is shrinking because of the cooling of the system.
Let the probability of a charm quark pair to be created at ${\rm \bf r}$ is $\rho(r)$, then survival
probability of charmonium in QGP medium becomes

\begin{equation}\label{Surv_Color}
S(p_{T}, R(N_{\rm part})) = \frac{\int_0^Rdr~r~\rho(r)~\phi(r,p_{T})}{\pi\int_0^Rdr~r~\rho(r)}.
\end{equation}

where $\phi$, the angle between ${\rm \bf p_{T}}$  and ${\rm \bf r}$, provides the range of escaping
possibility.
Integrating over $p_T$, Eq~(\ref{Surv_Color}) becomes the survival probability as a function of centrality

\begin{equation} 
  S(N_{\rm part})  = \int S(p_{T}, R(N_{\rm part}) ) \, dp_T.
\end{equation}
Where $R = R(N_{\rm part})$ is the medium size/radius, obtained in terms of the radius of the Pb nucleus given
by $R_0 = r_0\, A^{1/3}$ and the initial temperature of fireball, $T(N_{\rm part})$ created in each centrality
of the collisions is calculated as mentioned in the ref~\cite{URatioAbd}. 
With QGP formation time $\tau_{0}$ = 0.15 fm/$c$ at LHC~\cite{Initials1, Initials2}, we obtain
the initial temperature $T_{0}$ in most central collision is 0.65 GeV.

\renewcommand{\arraystretch}{1.4}
\begin{table}\label{prop}
\begin{center}
\caption{Charmonia properties from non-relativistic potential theory \cite{UPsi_Karsch,QPROP}.}
\begin{tabular}{c|c|c|c}
\hline
  {\rm Properties}        &    $J/\psi$   &  $\psi(2S)$      &  $\chi_c(1P)$ \\
\hline
 {\rm Mass~[GeV/$c^{2}$]}     &    3.1                    &    3.68                &   3.53 \\
\hline
{Radius \rm [fm]}          &    0.50                    &     0.90              &0.72    \\
\hline 
$\tau_{F}$ \rm [fm] \cite{UPsi_Karsch}  &    0.89        &     1.5              & 2.0      \\
\hline
$T_D$ \rm [GeV] used in work &    1.4~$T_C$            & 1.0~$T_C$       & 1.0~$T_C$    \\
\hline
\end{tabular}
\end{center}
\end{table}

\subsection{Medium dynamics at LHC}

The dynamics of central relativistic heavy ion collisions is modeled using the Bjorken
boost invariant picture with accelerated transverse expansion,  
resulting in a cylindrical volume in the geometry of the collision. 
Usually the system volume $V(\tau)$ is described as a system undergoing a isentropic expansion of
QGP fireball with the time-dependence of the volume $V(\tau)$ = $V_0 \tau/ \tau_0$ and the initial volume
$V_0$ = $\pi R^{2}(N_{\rm part}) \tau_0$~\cite{ThermalCharm_Zhang,Vol_FB}.
Introducing an acceleration term {\bf a}, the transverse radius
increases with proper time as 

\begin{equation}
  R(N_{\rm part},\tau) = R(N_{\rm part}) + a(\tau - \tau_0)^{2}/2
\end{equation}
    
with {\bf a} = 0.1 $c^{2}/fm$~\cite{ThermalCharm_Zhang}.
Now the expansion of fireball volume as function of radius and proper time is 

\begin{equation}
  V(\tau) = \pi R(N_{\rm part},\tau)^{2} (\tau{\bf p_{Z}}/M)
\end{equation}

\begin{figure}[htb]
\begin{center}
\begin{tabular}{cc}
  \includegraphics[width=0.58\textwidth]{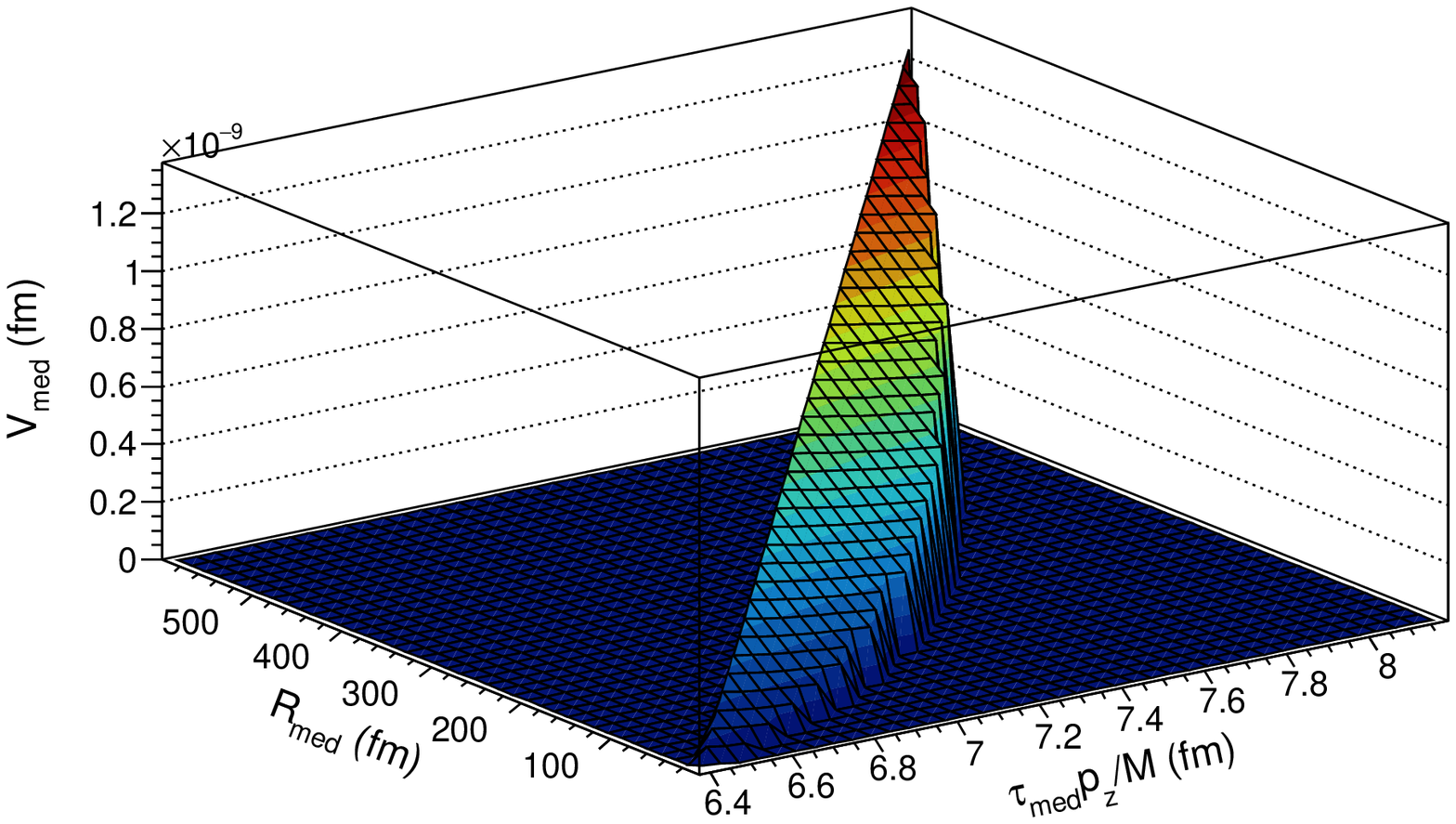}&
\includegraphics[width=0.40\textwidth]{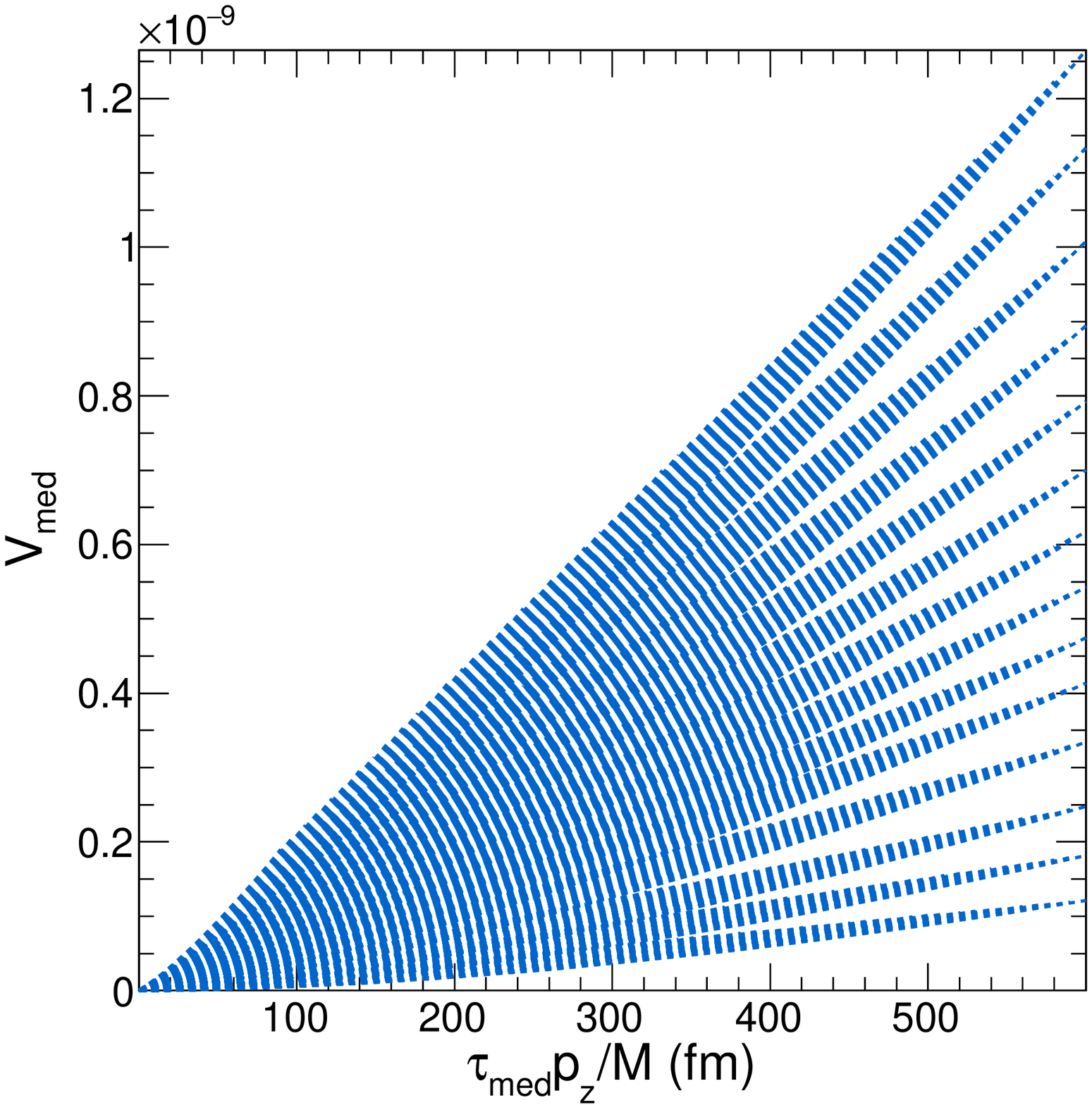}\\
\end{tabular}
\caption{The (Left) shows 3-Dimensional view of QGP fireball volume with transverse
radius $R(N_{\rm part},\tau)$ and with proper time of medium in the collision centrality 0-5\% and
(Right) shows the variation of the volume size in different centrality regions.}
\label{fig:Vol_QGP}
\end{center}
\end{figure}

The new term $({\bf p_{Z}}/M)$ is used to take into account the longitudinal momentum ($p_{Z}$)
of the charm-quark pairs, which ensures the longitudinal expansion of the volume during the QGP lifetime.
Thus we can have 2-dimensional (longitudinal+transverse) expansion of the fireball volume and
the temperature of the volume is decreasing as proper time $\tau$ increases as $T(\tau) = (\tau_{med}/\tau)^{1/3} T_0$.
The Fig~\ref{fig:Vol_QGP} (Left) shows 3D view of the fireball volume with transverse radius $R(N_{\rm part},\tau)$
and with proper time multiplied by ($p_{Z}/M$) in the collision centrality 0-5\% and (Right) shows the
variation of the volume size in different centrality regions.

\subsection{Recombination probability and formation rate}
The charmonium formation is happening in the deconfined medium through combination one of the $N_c$ charm
quarks with one of the $N_{\overline{c}}$ anti-charm quarks produced initially in a central heavy
ion collision. For a given charm quark, the probability $P$ to form a $\psi$ is
proportional to the number of available anti-charm quarks relative to the number of light anti-quark~\cite{Re_Prob}.

\begin{equation}
  P_{c \rightarrow \psi} \propto N_{\overline{c}}/N_{\overline{u} + \overline{d} + \overline{s}} \approx N_{c\overline{c}}/N_{ch} 
\end{equation}

We get the total number of $\psi$ expected in a given event by multiplying by the number of charm quarks $N_c$,

\begin{equation}
N_{\psi} \approx N_{c\overline{c}}^{2}/N_{ch} 
\end{equation}

We used the initial values $N_{c\overline{c}}$ = $N_{c}$ = $N_{\overline{c}}$.
From the above two equations we get the probability of charmonium formation in deconfinement medium. 

\begin{equation}\label{Prob_Reco}
N_{\psi}/N_{c\overline{c}} \approx N_{c\overline{c}}/N_{ch}  \approx P_{c \rightarrow \psi}
\end{equation}


\subsubsection{Kinetic model of formation and dissociation}

The recombination mechanism is the inverse process of thermal gluon dissociation of charmonium states,
that a free charm quark and anti-quark are captured in the $\psi$ bound state, emitting a color octet gluon.
It is to be noted that the recombination process is significant at low $p_T$, typically for values smaller
than the charmonium mass~\cite{Re_Prob,Re_Dis}.
The time evolution of charm quarks and charmonium states in the deconfined region, according to Boltzmann equation is

\begin{equation}\label{Bolt_Reco}
\frac{dN_{\psi}}{d\tau} = \Gamma_{F}N_{c}N_{\overline{c}}[V(\tau)]^{-1} - \Gamma_{D}N_{\psi}n_{g} ,
\end{equation}

where $n_g$ is the number density of gluons depending on the medium temperature. The widths $\Gamma_{F,D}$
are formation and dissociation reaction rates $ \langle \sigma v_{rel} \rangle$  averaged over the
momentum distribution of the participants ($c$ and $\overline{c}$ for $\Gamma_{F}$ and $\psi$ and $g$
for $\Gamma_{D}$) respectively~\cite{Re_Dis,Re_Prob,CS_Dis}.
There are analytical and numerical solutions for the equation Eq~(\ref{Bolt_Reco}). We use a different
approach to calculate the final number of charmonium and their survival probability in the medium. 

\subsubsection{Decoupling dissociation and recombination}
In this study, we consider the rates of the dissociation and recombination as two separate processes,
find their solutions and add them together to get the total number of survived charmonium states.
The dissociation and the recombination rates are given as

\begin{equation}\label{Decoup_Dis}
  \frac{dN_{\psi}^{D}}{d\tau} = - \Gamma_{D}N_{\psi}(0)~n_{g} 
\end{equation}

\begin{equation}\label{Decoup_Reco1}
\frac{dN_{\psi}^{F}}{d\tau} = \Gamma_{F}N_{c}N_{\overline{c}}[V(\tau)]^{-1} 
\end{equation}

For the gluon dissociation rate, the solution of Eq~(\ref{Decoup_Dis}) gives the number of charmonium states
survived the reaction. 

\begin{equation}\label{Decoup_Dis2}
  N_{\psi}^{D} = N_{\psi}(0)~exp^{-\int_{\tau_{0}}^{\tau_{f}} \Gamma_{D}n_{g}d\tau}
\end{equation}

The $N_{\psi}(0)$ (= $\sigma_{\psi}^{NN}~T_{AA}(\tau_0, b)$) is number of initially produced charmonia
in the collisions. $\sigma_{\psi}^{NN}$ is the production cross section in p+p collision and $T_{AA}$ is
nuclear overlap function taken from the ref~\cite{sigmaDir_Vogt} and~\cite{MCGlauber_dEnterria} respectively.
Before going to the solution of recombination rate, the probability that some of the newly formed $\psi$
could be dissociated by thermal gluons (although the rate will be very low initially), is to be taken
into account in the recombination process. As we assume $N_{\overline{c}} = N_{c} = N_{c\overline{c}}(Tot)$,
the Eq~(\ref{Decoup_Reco1}) becomes

\begin{equation}\label{Decoup_Reco2}
\frac{dN_{\psi}^{F}}{d\tau} = \Gamma_{F}N_{c\overline{c}}^{2}(Tot)[V(\tau)]^{-1} - \Gamma_{D}N_{\psi}n_{g}
\end{equation}

Where $N_{c\overline{c}}(Tot)$ is the sum of the $c\overline{c}$ pair produced in the initial collisions, $N_{c\overline{c}}(0)$
and those charm and anti-charm quarks separated in the dissociation of charmonium bound states $N_{c\overline{c}}^{Diss}$.
The $N_{c\overline{c}}(Tot)$ is

\begin{equation}\label{Ncc_Tot1}
  N_{c\overline{c}}(Tot) = N_{c\overline{c}}(0) + N_{c\overline{c}}^{Diss}
\end{equation}

with $N_{c\overline{c}}(0) = \sigma_{c\overline{c}}^{NN}~T_{AA}(\tau_0, b)$. Here $\sigma_{c\overline{c}}^{NN}$ is
the cross section for $c\overline{c}$ pair production in p+p collision~\cite{sigmCC_Thews55Tev}.
The new formation equation Eq~(\ref{Decoup_Reco2}) is analogous to that of radioactive decay chain reaction.
In the decay chain, the parent nucleus decays (here instead of decay, charmonium forms from two charm quarks,
then the number of charm quarks decreases as formation rate increases) to daughter nuclei which decays again
(here dissociate) to third nuclei. The solution of such differential equation can be found by Bateman equation
which take into account the effects of correlated mechanism of recombination from two charm quarks and
the dissociation of newly formed pairs. The solution of Eq~(\ref{Decoup_Reco2}) is then

\begin{equation}\label{Decoup_RecoTot}
 N_{\psi}^{F}  = \frac{\Lambda_{F}} {\Lambda_{D} - \Lambda_{F}}~N_{c\overline{c}}(Tot)
  [ e^{-\int_{\tau_{0}}^{\tau_{QGP}} \Gamma_{F}N_{c\overline{c}}^{2}(Tot)[V(\tau)]^{-1}d\tau}
      - e^{-\int_{\tau_{0}}^{\tau_{QGP}} \Gamma_{D}n_gd\tau} ]
    + N_{c\overline{c}}^{Diss}~e^{-\int_{\tau_{0}}^{\tau_{QGP}} \Gamma_{D}n_gd\tau}.   
\end{equation}
with $\Lambda_{F}$  =  $\int_{\tau_{0}}^{\tau_{QGP}} \Gamma_{F}N_{c\overline{c}}^{2}(Tot)[V(\tau)]^{-1}d\tau$ and
$\Lambda_{D}$  = $\int_{\tau_{0}}^{\tau_{QGP}} \Gamma_{D}n_gd\tau$.

Suppose there are $N_{\psi}(0)$ charmonium states initially at $\tau$ = 0 and each one has probability $Pr(\tau)$ to
dissociate in the time interval $\delta\tau$. With the dissociation rate $\Gamma_{D}$ (probability to
dissociate per unit time), the probability to dissociate is $Pr  = \Gamma_{D}n_{g}d\tau$.
Then the average number of $\psi$ that can be dissociated during the QGP lifetime is
$\int_{\tau_{0}}^{\tau_{QGP}} PrN_{\psi}(0)$ which is roughly equal to the number of charm quarks ($N_{c\overline{c}}^{Diss}$)
produced from the dissociated $\psi$ in the medium.

\begin{equation}\label{ProbDiss}
   \int_{\tau_{0}}^{\tau_{QGP}} PrN_{\psi}(0) =  N_{\psi}(0) \int_{\tau_{0}}^{\tau_{QGP}} \Gamma_{D}n_{g}d\tau  = N_{c\overline{c}}^{Diss}
\end{equation}

Now the Eq~(\ref{Ncc_Tot1}) becomes

\begin{equation}\label{Ncc_Tot2}
 N_{c\overline{c}}(Tot) = \sigma_{c\overline{c}}^{NN}~T_{AA}(\tau_0, b) + N_{\psi}(0) \int_{\tau_{0}}^{\tau_{QGP}} \Gamma_{D}n_{g}d\tau
\end{equation}

The number of recombined and survived $\psi$ is determined by the rates of dissociation
and recombination during the QGP lifetime. The solutions of these
differential rate equations are already found separately in earlier equations (Eq(\ref{Decoup_Dis2})
and (\ref{Decoup_RecoTot})). To get the total number of $\psi$ survived
at the end of QGP lifetime, we add the number of $\psi$ survived/recombined from the respective reactions.
The total number of $\psi$ survived the medium effect is 

\begin{align}\label{Decoup_TotalSol}
  N_{\psi}(\tau_{QGP})  &= \frac{\Lambda_{F}} {\Lambda_{D} - \Lambda_{F}}~N_{c\overline{c}}(Tot)
  [e^{-\int_{\tau_{0}}^{\tau_{QGP}} \Gamma_{F}N_{c\overline{c}}^{2}(Tot)[V(\tau)]^{-1}d\tau} 
    - e^{-\int_{\tau_{0}}^{\tau_{QGP}} \Gamma_{D}n_gd\tau}] \notag\\ 
  & + N_{c\overline{c}}^{Diss}~e^{-\int_{\tau_{0}}^{\tau_{QGP}} \Gamma_{D}n_gd\tau}  \notag\\ 
  & + N_{\psi}(0)~e^{-\int_{\tau_{0}}^{\tau_{QGP}} \Gamma_{D}n_gd\tau}.   
\end{align}

As mentioned in Eq~(\ref{Prob_Reco}), we can calculate the probability of the survival/recombination by
dividing the number of respective charmonium with sum of the initially produced $\psi$ and the total number
of charm-quarks pairs produced in the medium.
Therefore, we can have the survival probability of the $N_{\psi}^{D}$
(fractional of the survival of the gluon-dissociation) as 

\begin{equation}\label{Decoup_DisFrac}
  S(D) =  \frac{N_{\psi}^{D}} {N_{\psi}(0) + N_{c\overline{c}}(Tot)} = \frac{N_{\psi}(0)} {N_{\psi}(0) + N_{c\overline{c}}(Tot)}
  exp^{-\int_{\tau_{0}}^{\tau_{QGP}} \Gamma_{D}n_{g}d\tau} 
\end{equation}

Similarly the probability of recombination (fractional of the formation/recombination) in the medium is

\begin{align}\label{Decoup_RecoFrac}
  S(F) & = \frac{N_{\psi}^{F}}{N_{\psi}(0) + N_{c\overline{c}}(Tot)} =  \notag\\
  & \frac{N_{c\overline{c}}(Tot)}{N_{\psi}(0) + N_{c\overline{c}}(Tot)} \frac{\Lambda_{F}} {\Lambda_{D} - \Lambda_{F}}
  [ e^{-\int_{\tau_{0}}^{\tau_{QGP}} \Gamma_{F}N_{c\overline{c}}^{2}(Tot)[V(\tau)]^{-1}d\tau} - e^{-\int_{\tau_{0}}^{\tau_{QGP}} \Gamma_{D}n_gd\tau} ]  \notag\\ 
   & + \frac{N_{c\overline{c}}^{Diss}}{N_{\psi}(0) + N_{c\overline{c}}(Tot)}~e^{-\int_{\tau_{0}}^{\tau_{QGP}} \Gamma_{D}n_gd\tau}
\end{align}

Since the $N_{c\overline{c}}^{Diss}$ is very small number compared to $N_{c\overline{c}}(0)$, the last term of the above equation
can be omitted in the final calculation of survival probability.

The total survival probability of the charmonium in the medium is the product of Eq~(\ref{Surv_Color}),
(\ref{Decoup_DisFrac}) and (\ref{Decoup_RecoFrac}).

\begin{align}\label{Surv_Tot}
  S(p_{T}, R(N_{part})) &= \frac{1}{N_{\psi}(0) + N_{c\overline{c}}(Tot)} \int_{0}^{R}dr~r~\rho(r)~\phi(r,p_{T})  \notag\\ 
  & (\frac{\Lambda_{F}} {\Lambda_{D} - \Lambda_{F}}  N_{c\overline{c}}(Tot)
  [ e^{-\int_{\tau_{0}}^{\tau_{QGP}} \Gamma_{F}N_{c\overline{c}}^{2}(Tot)[V(\tau)]^{-1}d\tau} - e^{-\int_{\tau_{0}}^{\tau_{QGP}} \Gamma_{D}n_gd\tau} ])  \notag\\ 
   &~N_{\psi}(0)e^{-\int_{\tau_{0}}^{\tau_{QGP}} \Gamma_{D}n_{g}d\tau} \notag\\ 
  \end{align}


  The nuclear modification factor, $R_{AA}$ is obtained from survival probability taking into account 
the feed-down corrections as follows,

\begin{eqnarray}
 R_{\rm AA}(\chi_c(1P)) &=& S(\chi_{c1} + \chi_{c2})  \nonumber\\
 R_{\rm AA}(\psi(2S)) &=& S(2S) \nonumber\\
 R_{\rm AA}(\psi(1S)) &=& g_1 ~S(1S) + g_2~S(1P) + g_3~S(2S) \nonumber\\
\end{eqnarray}
The factors $g$'s are obtained from the measurement in proton-nucleon and pion-nucleon
interactions at 300 $GeV$~\cite{Digal_feedDown}. The values of $g_{1}$, $g_{2}$ and $g_{3}$ are
0.62, 0.3 and 0.08 respectively. 

\section{Results and discussions}

\begin{figure}[htb]
\begin{center}
\begin{tabular}{cc}
\includegraphics[width=0.48\textwidth]{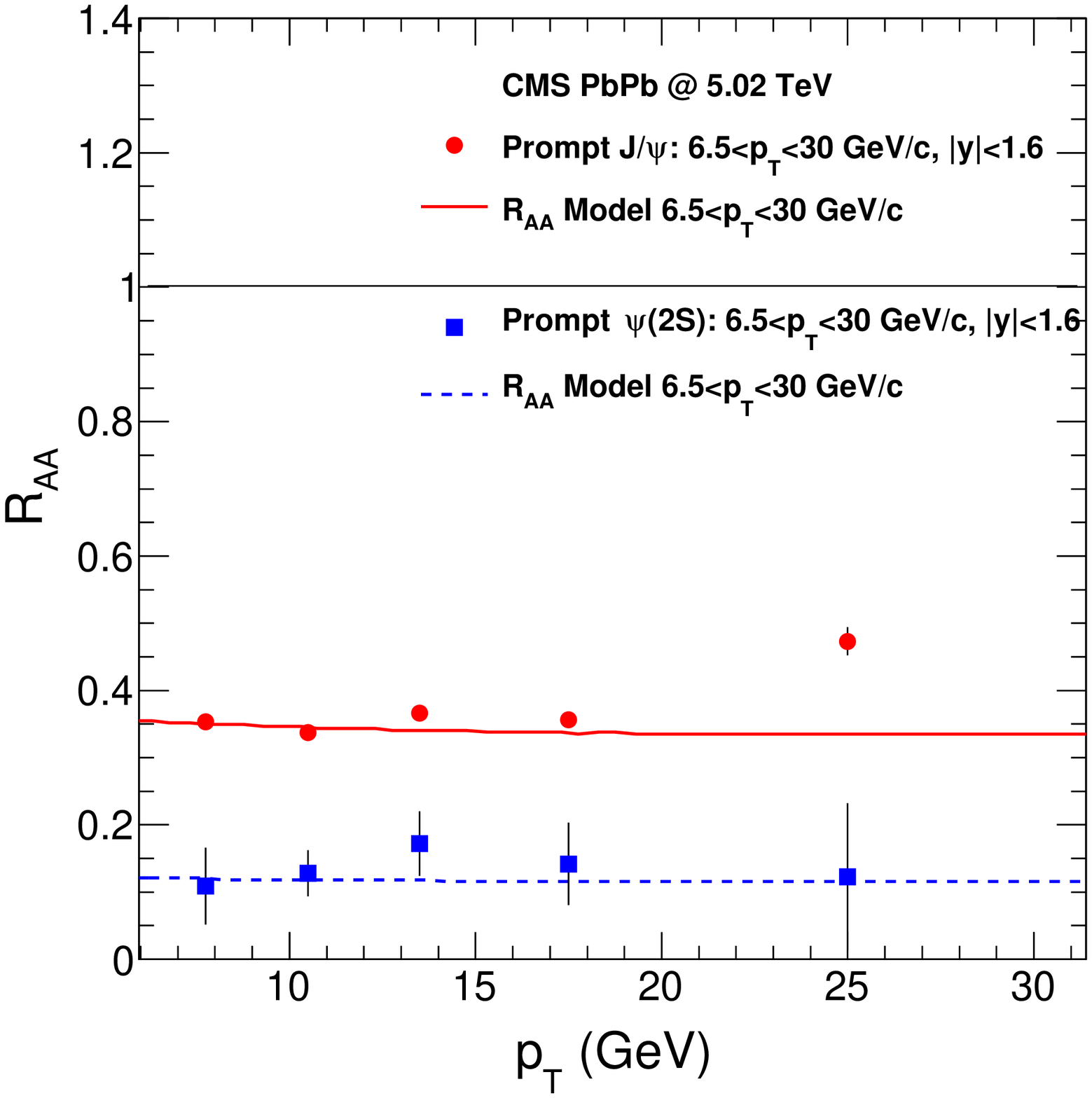}&
\includegraphics[width=0.48\textwidth]{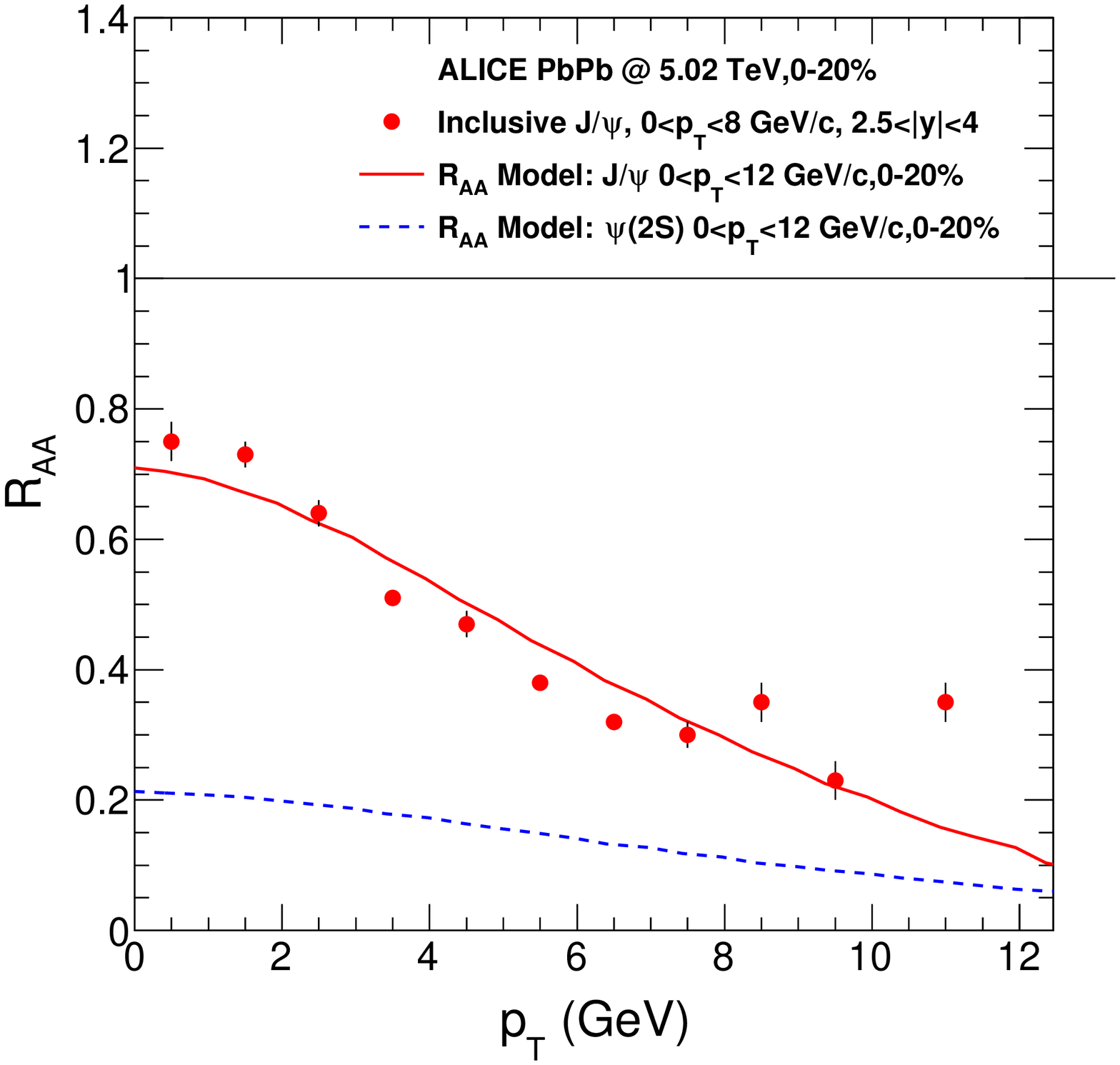}\\
\end{tabular}
\caption{(a) The nuclear modification factor, $R_{AA}$
as a function of $p_{T}$ for $J/\psi$ and $\psi(2S)$. The solid points are measured $R_{AA}$ by
CMS experiment in Pb+Pb collisions at $\sqrt{s_{\rm NN}}$ = 5.02 TeV. (b) The nuclear
modification factor, $R_{AA}$ as a function of $p_{T}$ for $J/\psi$ measured with ALICE experiment.
The solid and dashed lines in all figures represent the model calculations.}
\label{fig:JpsiRaaCMSALICE}
\end{center}
\end{figure}

With the model calculation, we calculated the nuclear modification factor,
$R_{AA}$ of $J/\psi$ and $\psi(2S)$ as function of $p_{T}$ and centralities relevant for LHC experiments.
The calculations are compared with the data measured at CMS and ALICE Experiments.
The survival probabilities of resonance states has a unique $p_T$ dependence decided by the $T_D$,
$\tau_{F}$, $T(\tau)$ and $\tau_{med}$ for each $\psi$ state. The $R_{AA}$
of $J/\psi$ and $\psi(2S)$ as a function of $p_{T}$ is shown in Figure~\ref{fig:JpsiRaaCMSALICE} (Left).
The solid circles and squares are the measured $R_{AA}$ for $J/\psi$ and $\psi(2S)$ respectively in high $p_{T}$
(6.5-30.0 Gev/c) and mid rapidity region, with CMS experiment in Pb+Pb collisions at $\sqrt{s_{\rm NN}}$
= 5.02 TeV \cite{JCMS502TeV}. Similarly the $R_{AA}$ as a function of $p_{T}$ (0-12 GeV/c) is shown in
Figure~\ref{fig:JpsiRaaCMSALICE} (Right). The solid and dashed lines are the model calculations for $R_{AA}$ in the
respective $p_{T}$ regions. The interplay between different medium-induced reactions decides the trend of
$p_{T}$ curve in all regions. The model replicates the trend of the $p_T$ dependence of the measured $R_{\rm AA}$
except in the last bin of $J/\psi$ high $p_{T}$ region. This may be because of less energy loss of high $p_{T}$ charmonia
as predicted in an energy loss model~\cite{Fransua}.

\begin{figure}[htb]
\begin{center}
\includegraphics[width=0.48\textwidth]{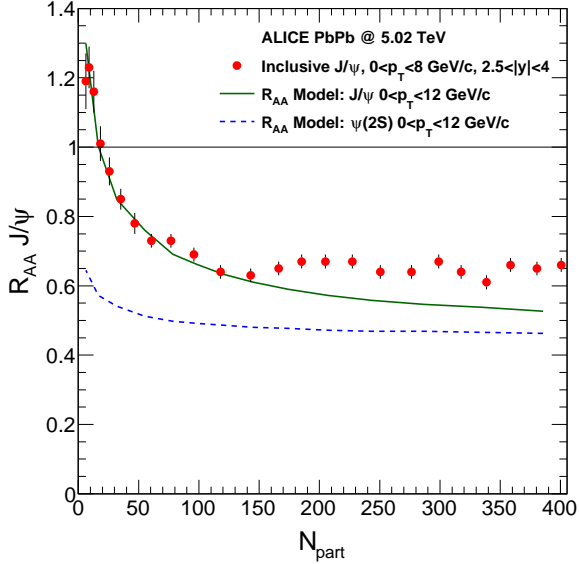} 
\caption{The nuclear modification factor, $R_{AA}$
as a function of $N_{\rm part}$ for $J/\psi$. The solid points are the measured $R_{AA}$ by
ALICE experiment in Pb+Pb collisions at $\sqrt{s_{\rm NN}}$ = 5.02 TeV \cite{ALICE502TeV_Psi}.
The lines represent the present model calculations.} 
\label{fig:RaaALICE0pT80}
\end{center}
\end{figure}

\begin{figure}[htb]
\begin{center}
\begin{tabular}{cc}
\includegraphics[width=0.48\textwidth]{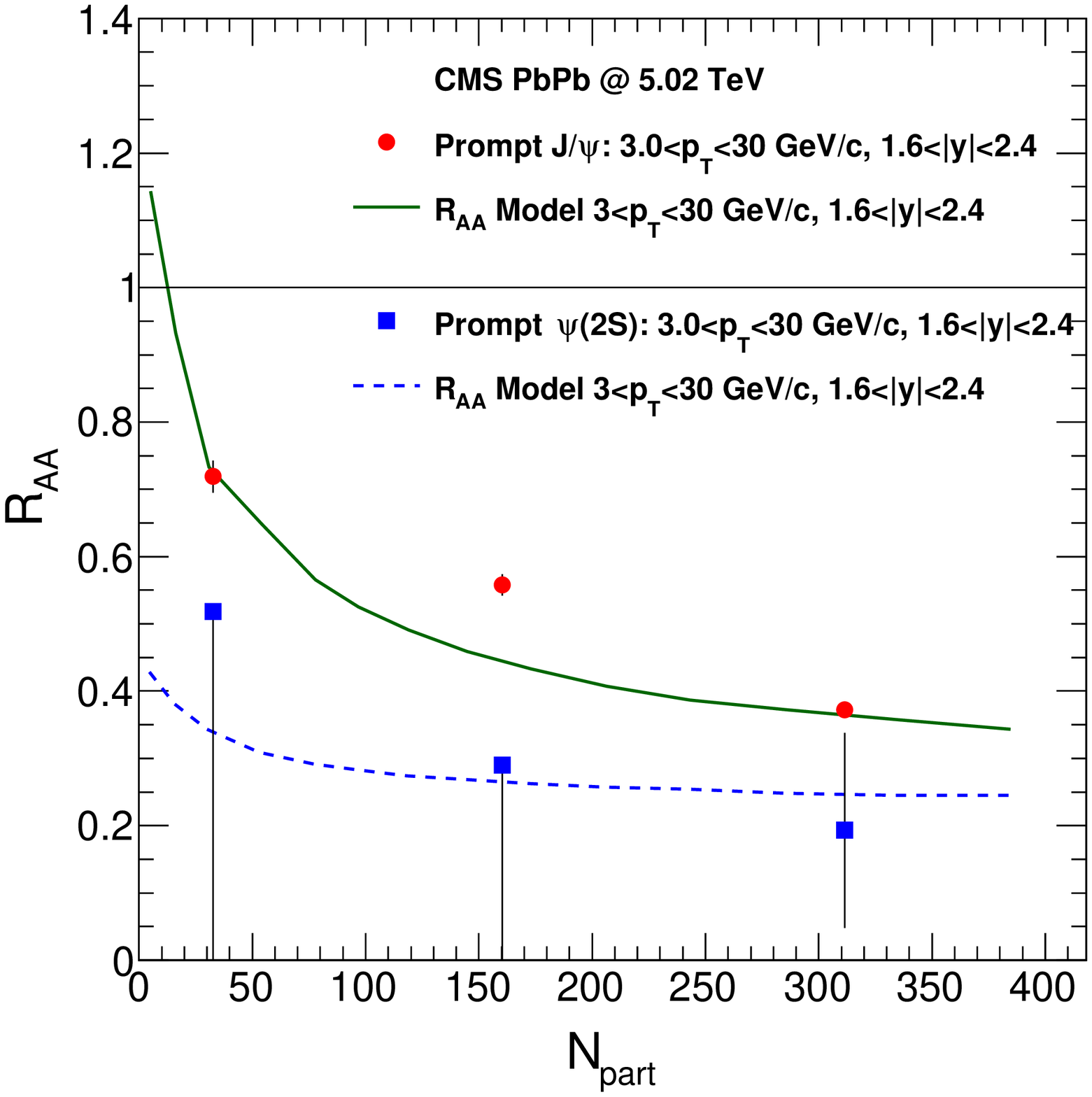} &
\includegraphics[width=0.48\textwidth]{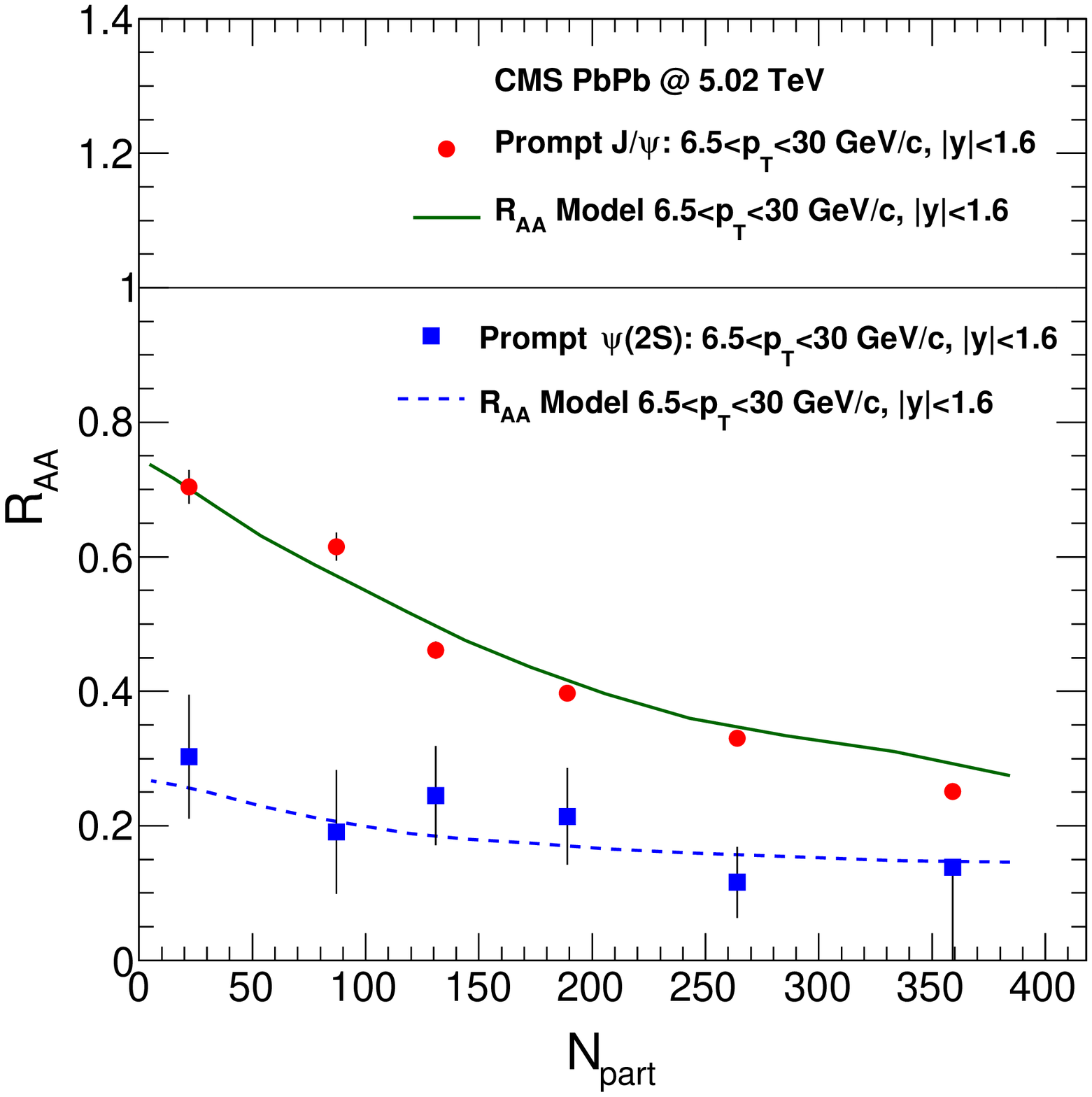} \\
\end{tabular}
\caption{(Left) The nuclear modification factor, $R_{AA}$
  as a function of $N_{\rm part}$ for $J/\psi$ and $\psi(2S)$ with low $p_{T}$ (3-30 GeV/c) and forward rapidity.
  The solid circles and squares are the measured $R_{AA}$ of $J/\psi$ and $\psi(2S)$ respectively with CMS experiment
  in Pb+Pb collisions at $\sqrt{s_{\rm NN}}$ = 5.02 TeV \cite{JCMS502TeV} (Right) Same as in Left with high $p_{T}$
  (6.5-30 GeV/c) and mid rapidity. The lines represent the present model calculations.}
\label{fig:JpsiCMSRaa}
\end{center}
\end{figure}

Figure~\ref{fig:RaaALICE0pT80} shows the $R_{AA}$ as a function of $N_{\rm part}$ for $J/\psi$ 
and $\psi(2S)$ with $p_{T}$ $<$ 12.0 GeV/c. The solid red circles are $R_{AA}$ data measured by ALICE experiment in
Pb+Pb collisions at $\sqrt{s_{\rm NN}}$ = 5.02 TeV at forward rapidity and $p_{T}$ $<$ 8.0 GeV/c~\cite{ALICE502TeV_Psi}.
The solid line, the present model calculation agrees well with the measured data keeping in mind that the
measured $R_{AA}$ is for inclusive $J/\psi$ while the model calculation is for prompt $J/\psi$ and $\psi(2S)$.
Figure~\ref{fig:JpsiCMSRaa} shows the the nuclear modification factor, $R_{AA}$ as a function of $N_{\rm part}$
for $J/\psi$ and $\psi(2S)$ with low $p_{T}$ (3-30 GeV/c) and forward rapidity (Left) and with high $p_{T}$
(6.5-30 GeV/c) and mid rapidity (Right). The lines in both figures are representing the model calculations.
The model reproduces well the measured nuclear modification factors of both J/$\psi$ and $\psi(2S)$
in all centralities.

The suppression of resonance states increases with increasing centrality (increasing $N_{\rm part}$)
as expected. Also the $\psi(2S)$ is more suppressed than J/$\psi$ matching with the scenario of sequential
suppression in the measured $R_{AA}$~\cite{JCMS,JCMS502TeV}. The calculated suppression is the combined result
of color screening, gluon-dissociation and recombination reactions. In Figure~\ref{fig:RaaALICE0pT80},
the suppression increases steeply up-to $N_{\rm part}$ = 100 and then it becomes a slow suppression indicating
the overplay of recombination reaction in lower $p_{T}$ region. To note that the inclusive $J/\psi$ consists
of prompt charmonium (directly produced from parton collisions and feed-down contribution) and non-prompt
charmonium (decayed from  the B-meson). At higher $p_{T}$ region, the calculation shows a
smooth suppression plotted as in Figure~\ref{fig:JpsiCMSRaa}. Since the medium effects are not significant
in peripheral collision (lower $N_{\rm part}$), the $R_{AA}$ values
go beyond the unity as shown in Figure~\ref{fig:RaaALICE0pT80} and \ref{fig:JpsiCMSRaa}.
There can be other sources which may contribute in the suppression of charmonia states, like suppression
due to initial nuclear suppression (shadowing effect) which we assume to be counted in the amount of
color screening effect/dissociation and hence are not calculated separately in the present work.
The important parameters of the model like formation time ($\tau_F$), radius and dissociation temperatures
($T_D$) are obtained from temperature-dependent potential models which reproduce the quarkonia spectroscopy
very well~\cite{UPsi_Mocsy2}.

\section{Conclusions}
We calculate the nuclear modification factors of charmonia states ($J/\psi$ and $\psi(2S)$) in an expanding QGP
of finite lifetime and size produced in Pb+Pb collisions at $\sqrt{s_{NN}}=$ 5.02 TeV. The nuclear modification is
resulted from the combined effect of color screening, gluon-dissociation and recombination reactions. The competition
between the resonance formation time $\tau_{F}$, the medium temperature $T(\tau)$, $\tau_{QGP}$ and etc decide
the dependence of the survival probabilities of $\psi$ states in different kinematic regions.
The dynamics of central relativistic heavy ion collisions is modeled using the Bjorken
boost invariant picture with accelerated transverse expansion resulting in a cylindrical volume of fireball.
The calculated suppressions are compared with the $R_{AA}$ measured at CMS and ALICE Experiments. The model
reproduces well the measured nuclear modification factors of both J/$\psi$ and $\psi(2S)$ in most of the centrality
regions. In addition, the scenario of sequential suppression in the measured $R_{AA}$ at LHC experiments are reflected
well in the model calculations.

\section{Acknowledgment}
This Project was funded by the Deanship of Scientific Research (DSR) at King Abdulaziz University, Jeddah, under
grant no. G: 589-130-1439. The authors, therefore acknowledge with thanks DSR for technical and financial support.

\end{document}